\documentclass[10pt,conference, letterpaper]{IEEEtran}
\IEEEoverridecommandlockouts

\usepackage{configuration}
\input{acro.cfg}

\begin{document}
\bstctlcite{IEEEexample:BSTcontrol}


\title{Conflict Detection in AI-RAN:
Efficient Interaction Learning and Autonomous Graph Reconstruction}

\author{
  \IEEEauthorblockN{
    Joao F. Santos\IEEEauthorrefmark{1},
    Arshia Zolghadr\IEEEauthorrefmark{1}, 
    Scott Kuzdeba\IEEEauthorrefmark{2}, and 
    Jacek Kibi\l{}da\IEEEauthorrefmark{1}
  }
  \IEEEauthorblockA{
    \IEEEauthorrefmark{1}\textit{Commonwealth Cyber Initiative},
    \textit{Virginia Tech}, USA,
    e-mail: \{joaosantos, arshiaz, jkibilda\}@vt.edu}
    \IEEEauthorrefmark{2}\textit{BAE Systems, Inc},
    e-mail: scott.kuzdeba@baesystems.us

}
\maketitle

\begin{abstract}

  \ac{AI}-native mobile networks represent a fundamental step toward 6G, where
learning, inference, and decision making are embedded into the \ac{RAN} itself. 
In such networks, multiple \ac{AI} agents optimize the network to
achieve distinct and often competing objectives. As such, conflicts become 
inevitable and have the potential to degrade performance, cause instability, 
and disrupt service. Current approaches for conflict detection rely
on conflict graphs created from relationships between \ac{AI} agents, 
parameters, and \acp{KPI}.
Existing works often rely on complex and computationally expensive 
\acp{GNN} and depend on manually chosen thresholds to create conflict
graphs. 
In this work, we present the first systematic framework for conflict detection 
in \ac{AI}-native mobile networks, propose an efficient
two-tower encoder architecture for learning interactions based on data from 
the \ac{RAN}, and introduce a data-driven sparsity-based mechanism for 
autonomously reconstructing conflict graphs without manual fine-tuning. 
\end{abstract}



\acresetall

\iftrue
\fancypagestyle{firstpage}
{
    \fancyhead[L]{This work has been submitted to the IEEE for possible
      publication.\\
      Copyright may be transferred without notice, after which this version may no longer be accessible.}
    \fancyhead[R]{}
    \pagenumbering{gobble}
}

\thispagestyle{firstpage}

\fi

\section{Introduction}\label{sec:intro}

One of the major developments toward 6G is the shift toward \ac{AI}-native
mobile networks, where learning, inference, and decision-making are no longer
treated as external add-ons, but are instead integrated into the network
architecture itself~\cite{khan2023ai}. Rather\,than\,relying on static,
manually configured policies, the \ac{RAN} autonomously learns from data, makes 
real-time control decisions, and continuously adapts to network conditions, e.g., 
traffic demand, user mobility, and channel state~\cite{kundu2025ai}. 
Such autonomy is achieved through a collection of intelligent controllers, e.g., 
dApps, xApps, and rApps in the context of O-RAN~\cite{santos2025managing}, 
or more broadly, through \ac{AI} agents that perform 
learning-enabled control loops for optimizing the network to cater to 
diverse use cases and applications~\cite{khan2023ai, mallu2023ai, baron2024eavesdropper}.

As multiple \ac{AI} agents operate concurrently on the mobile network 
infrastructure, they may attempt to execute actions to achieve distinct 
or even conflicting objectives. 
For example, an agent may attempt to increase the transmit power to maximize
throughput, while another agent simultaneously attempts to decrease the transmit
power to minimize energy consumption~\cite{del2025pacifista}. 
This constitutes a \textit{direct conflict}, which can degrade performance, 
cause instability, or disrupt service~\cite{del2025pacifista}. Other 
types of conflicts also exist, including \textit{indirect} and 
\textit{implicit} conflicts. For formal definitions and examples, we refer the reader 
to~\cite{oran_conflict_mitigation_2024, zolghadr2025learning}.
Despite the potential vulnerabilities introduced by conflicting 
control decisions in \ac{AI}-native mobile networks, research on 
conflict detection and mitigation remains in its early stages.

\begin{figure}[t]
    \centering
    \includegraphics[width=0.90\columnwidth]{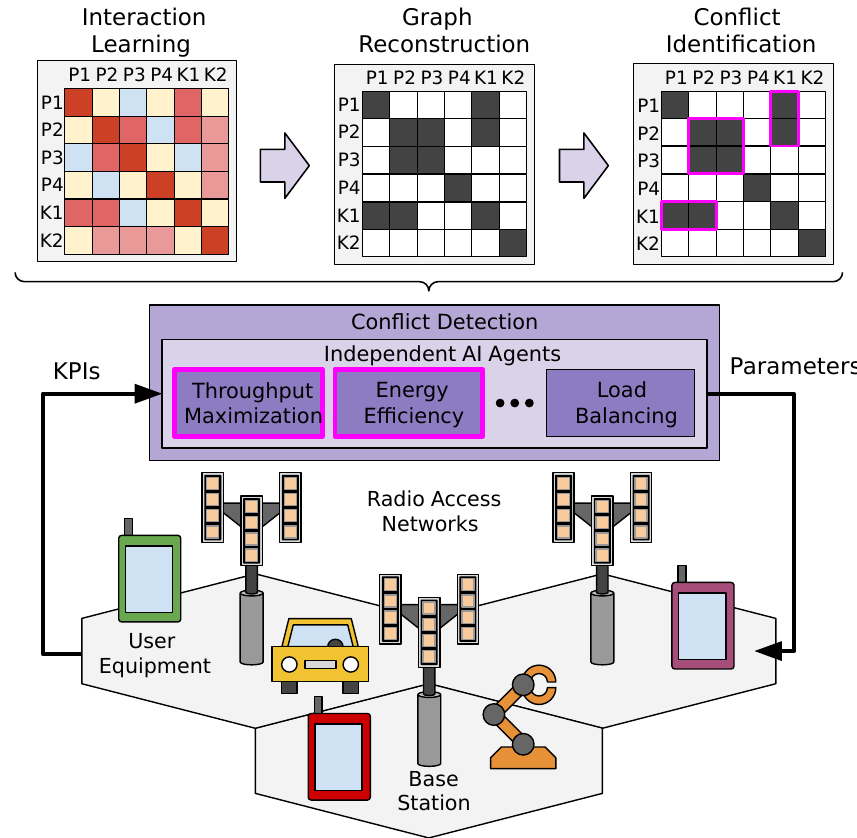}
    \vspace{-0.5em}
    \caption{Our proposed conflict detection framework for \ac{AI}-native
    mobile networks, where independent \ac{AI} agents control
    the\,\ac{RAN}\,to\,achieve\,distinct and possibly conflicting objectives.}
    \label{fig:intro}
\end{figure}

In the current literature, the relationships between \ac{AI} agents, 
the \ac{RAN} parameters they control, and the \acp{KPI} they subscribe 
to are represented as graphs~\cite{del2025pacifista, zolghadr2025learning,
sharma2025towards}, and conflicts are detected by analyzing 
the topology of these graphs~\cite{del2025pacifista}. 
The first challenge in this paradigm is inferring the 
graph topology from data collected from the \ac{RAN}, which is achieved 
by learning interactions between agents, parameters, and \acp{KPI}.
To date, autonomous methods such as \acp{GNN} have been 
proposed~\cite{zolghadr2025learning}, at the cost of complex and 
computationally expensive message-passing operations~\cite{shao2024distributed}. 
Once interactions are learned, an appropriate decision rule is applied 
to determine which interactions are meaningful and constitute
relationships to create the graph topology.
Here, existing works rely on manually chosen thresholds to identify 
relevant interactions~\cite{wadud2024qacm, del2025pacifista}, an approach 
that is difficult to generalize across datasets and deployment scenarios. 
Furthermore, while existing works address different aspects of the problem of
conflict detection in mobile networks, there is a lack of a systematic 
framework to 
characterize their design choices and trade-offs.

In this work, we address these open challenges by first framing 
conflict detection in \ac{AI}-native mobile networks 
as a three-stage problem comprising \1 interaction
learning, \2 graph reconstruction, and \3 conflict identification,
as shown in Fig.~\ref{fig:intro}.
Next, we posit interaction learning as a ranking problem, often encountered in 
recommender systems~\cite{yan2024ranktower},
and propose a lightweight 
two-tower encoder architecture for efficiently learning interactions based on
data  collected from the \ac{RAN}~\cite{li2022inttower}. 
Then, we develop a data-driven, sparsity-based mechanism that
autonomously selects relevant interactions~\cite{martins2016softmax}, 
enabling conflict graph reconstruction without manual fine-tuning across 
datasets and scenarios.
Finally, we evaluate our conflict detection solution using the rule-based 
conflict identification approach proposed in~\cite{zolghadr2025learning}.

The main contributions of this paper are as follows:

\begin{itemize}[leftmargin=1em]

  \item We propose a framework for decomposing 
    conflict detection 
    as a three-stage problem 
    comprised of \1 interaction learning, \2 graph reconstruction, and 
    \3 conflict identification.
  \item We model the interaction learning stage as
    a ranking problem between parameters and \acp{KPI}, and propose a 
    two-tower encoder architecture for efficiently learning interactions.
  \item We introduce a data-driven, sparsity-based approach for selecting relevant
    interactions, enabling autonomous graph reconstruction 
    without the need for manual fine-tuning.
  \item We validate the efficiency of our proposed conflict detection solution 
    using a dataset of Gaussian-distributed parameters and \acp{KPI} generated 
    from a known conflict model~\cite{banerjee2021toward}. 

\end{itemize}

In the following sections, we present our systematic framework for conflict 
detection, and assess related works with respect to their design choices. 
Next, we introduce our proposed solutions for efficient interaction 
learning and autonomous graph reconstruction, and compare their performance 
against existing solutions. Finally, we pose our concluding remarks. 

\section{The Three Stages of Conflict Detection}\label{sec:decomp}

The existing literature on conflict detection in \ac{AI}-native mobile networks
commonly adopts a graph-based model~\cite{del2025pacifista, wadud2024qacm,
sharma2025towards} to represent \ac{AI} agents, 
\ac{RAN} parameters, and \acp{KPI} as nodes, and their relationships as edges,
forming a conflict graph \(\mathcal{G} = (\mathcal{V}, \mathcal{E})\). 
We can denote the set of independent \ac{AI} agents
as $\mathcal{A}$, the set of \ac{RAN} parameters as $\mathcal{P}$, and
the set of \acp{KPI} as $\mathcal{K}$, and define the set of vertices  
$\mathcal{V} = \mathcal{A} \cup \mathcal{P} \cup \mathcal{K}$. The set of edges
\(\mathcal{E}\) describes the relationships between 
these entities, whereby an edge between \1 an agent \(a\!\in\!\mathcal{A}\) and a parameter
\(p\!\in\!\mathcal{P}\) represents that the agent controls the parameter, 
\2 a parameter \(p\!\in\!\mathcal{P}\) and a \ac{KPI} 
\(k\!\in\!\mathcal{K}\) represents that the parameter influences the KPI, 
and \3 a \ac{KPI} \(k\!\in\!\mathcal{K}\) and an agent \(a\!\in\!\mathcal{A}\) 
represents that the value of the \ac{KPI} informs the agent's decision. 
This modeling approach enables the identification and categorization of conflicts 
through the analysis of the graph structure, represented by an~adjacency~matrix~\(\bm{A}\). 

The relationships between agents and the parameters they control,
as well as the \acp{KPI} they observe to make decisions, are 
often explicit and known by design, 
since agents must request access to the network infrastructure for 
modifying control parameters and accessing 
\acp{KPI}~\cite{santos2025managing}. In contrast, the
relationships between parameters and \acp{KPI} are often non-trivial and
dynamic, emerging over time or manifesting under specific operating
conditions~\cite{zolghadr2025learning, sharma2025towards}. 
When combined with interdependencies among parameters themselves, and  
among \acp{KPI}, these relationships can form complex chains of dependencies 
that are difficult to identify in advance. As a result, the
major challenge for conflict detection in \ac{AI}-native mobile networks 
lies in learning relationships between parameters and \acp{KPI},
while also capturing relationships within each of these entities, 
based on data collected from the \ac{RAN}.
This problem can be decomposed into three stages: interaction learning, graph reconstruction, and conflict identification.

To formalize this decomposition, we consider datasets of samples 
collected from the \ac{RAN}, e.g., from simulations or real-world observations.
Let \(\bm{X}_p \! \in \! \mathbb{R}^{N_p \times L}\) denote $L$ samples
of \(N_p= \! |\mathcal{P}|\) parameters, and 
\(\bm{X}_k \!\in \! \mathbb{R}^{N_k \times L}\) denote
$L$ samples of $N_k \! =\! |\mathcal{K}|$ \acp{KPI}.
While, a binary matrix
$\bm{A}_{\texttt{known}} \! \in \! \{0,1\}^{N_a \times (N_p + N_k)}$
encodes the known relationships between the $N_a =
|\mathcal{A}|$ agents with parameters and \acp{KPI} based on the information 
obtained from the \ac{RAN} control plane (in the case of O-RAN, these would 
be obtained through the Subscription Manager~\cite{santos2025managing}).
An element $[\bm{A}_{\texttt{known}}]_{i,j}\! = \! 1$ 
indicates that agent $i$ controls (or subscribes) to parameter (or \ac{KPI}) $j$, 
and $[\bm{A}_{\texttt{known}}]_{i,j}\! =\! 0$ indicates otherwise.  
Then, we define the following three stages of conflict detection: 

\begin{itemize}[leftmargin=1em]

  \item{\textbf{Stage I -- Interaction Learning}}: This stage focuses
    on \emph{learning interactions} between parameters and \acp{KPI}
    from data collected from the \ac{RAN}. Given \(\bm{X}_p\) and \(\bm{X}_k\), a
    reconstruction approach assigns \textit{scores} capturing the
    relevance of their interactions:
    \begin{equation}
      \bm{S} = \mathcal{R}(\bm{X}_p, \bm{X}_k) =
\begin{bmatrix}
\bm{S}_{pp} & \bm{S}_{pk} \\
\bm{S}_{kp} & \bm{S}_{kk}
\end{bmatrix},
\label{eq:score_matrix}
    \end{equation}

    where \(\mathcal{R}(\cdot)\) denotes the learning method, 
    e.g., statistical profiling, causal inference, or \acp{GNN}.
    The chosen learning method dictates the type of score, e.g.,
    Kolmogorov-Smirnov distance~\cite{del2025pacifista},
    correlation~\cite{zolghadr2025learning}, or similarity, as well as the 
    structural properties of the resulting score matrix 
    \(\bm{S} \in \mathbb{R}^{(N_p+N_k) \times (N_p+N_k)}\).
    This matrix contains blocks capturing 
    cross-entity interactions between 
    parameters and \acp{KPI} ($\bm{S}_{pk}\!\in\!\mathbb{R}^{N_p \times N_k}$), 
    as well as same-entity interactions among parameters 
    ($\bm{S}_{pp}\!\in\!\mathbb{R}^{N_p \times N_p}$) 
    and among \acp{KPI} ($\bm{S}_{kk}\!\in\!\mathbb{R}^{N_k \times N_k}$).
    Symmetric learning methods, e.g., statistical profiling and \acp{GNN}, yield 
    $\bm{S}_{kp} = \bm{S}_{pk}^\top$, whereas causal inference methods may
    induce asymmetry, i.e., $\bm{S}_{kp} \neq \bm{S}_{pk}^\top$.

  \item{\textbf{Stage II -- Graph Reconstruction}}: The second stage focuses on 
    \textit{re-creating conflict graphs} from learned scores and 
    known relationships from agents. 
    Given $\bm{S}$, a binarization approach determines the relevant interactions and  
    maps the real-valued scores into a partial adjacency matrix: 
    \begin{equation}
      \hat{\bm{A}}_{\texttt{learned}} = \mathcal{B}(\bm{S}),
    \end{equation}

    where \(\mathcal{B}(\cdot)\) denotes the score binarization method, e.g.,
    thresholds~\cite{del2025pacifista}, sparsity, or other strategies.  
    $\hat{\bm{A}}_{\texttt{learned}} \in \{0,1\}^{(N_p+N_k)\times (N_p+N_k)}$
    encodes topological information describing relationships between parameters 
    and \acp{KPI}, as well as among parameters, and among \acp{KPI}.
    We augment  $\hat{\bm{A}}_{\texttt{learned}}$ with $\bm{A}_{\texttt{known}}$ 
    to generate the adjacency matrix~\cite{del2025pacifista}:
    \begin{equation}
      \label{eq:box}
        \hat{\bm{A}} = \hat{\bm{A}}_{\texttt{learned}} \boxplus \bm{A}_{\texttt{known}} = 
        \begin{bmatrix}
\bm{I}_{N_a} & \bm{A}_{\texttt{known}} \\
\bm{A}_{\texttt{known}}^T & \hat{\bm{A}}_{\texttt{learned}}
        \end{bmatrix},
    \end{equation}
    
    where $\boxplus$ denotes a block-wise matrix augmentation operation, 
    adding $\bm{A}_{\texttt{known}}$ as additional 
    rows above $\bm{A}_{\texttt{learned}}$, along with 
    the corresponding $\bm{A}_{\texttt{known}}^\top$ as additional columns before 
    $\bm{A}_{\texttt{learned}}$~\cite{del2025pacifista}, and filling the 
    remaining block corresponding to relationships among agents 
    as an identity matrix $\bm{I}_{N_a}$. This operation yields a 
    binary adjacency matrix 
     $\hat{\bm{A}} \in \{0,1\}^{(N_a+N_p+N_k)\times (N_a+N_p+N_k)}$ that 
     approximates $\bm{A}$ and represents a conflict graph.
    
  \item{\textbf{Stage III -- Conflict Identification}}: 
    The final stage focuses on analyzing the graph topology to identify
    different types of conflicts, i.e., direct, indirect, and
    implicit~\cite{oran_conflict_mitigation_2024}. Given $\hat{\bm{A}}$, an identification 
    approach searches the conflict graph for subgraphs representing the different
    conflicts:  
    \begin{equation}
      C = \mathcal{I}(\hat{\bm{A}}),
    \end{equation}

    where \(\mathcal{I}(\cdot)\) denotes the conflict identification method,
    e.g., rule-based~\cite{zolghadr2025learning}, probabilistic, or exact graph
    isomorphism algorithms~\cite{carletti2015vf2}. 
    The resulting
    set \(\mathcal{C}\) contains the different types of identified conflicts between 
    $\mathcal{A}$, $\mathcal{P}$ and $\mathcal{K}$. 

\end{itemize}

We can now express the conflict detection problem as a combination of the
three stages, as illustrated in Fig.~\ref{fig:intro}:
\begin{equation}
C = \mathcal{I}\big(\mathcal{B}(\mathcal{R}(\bm{X}_p, \bm{X}_k)) \boxplus \bm{A}_{\texttt{known}}\big).
\end{equation}

In the context of O-RAN, this systematic framework for conflict detection
can be realized as one or more rApps, or as a native conflict detection module
within the \ac{Non-RT RIC}, leveraging the O1 interface to
collect the telemetry data from the \ac{RAN} and the inference
platform service from the AI/ML Framework~\cite{lee2021ran} to learn interactions 
and reconstruct conflict graphs, and subsequently identify 
conflicts. In addition, its placement in the \ac{Non-RT RIC} would make it well 
positioned to \1 raise alarms to inform the network operator of potential 
conflicts, and/or \2 automatically generate policies to trigger conflict 
mitigation solutions.
For information about the O-RAN architecture, components and interfaces, 
we refer the reader to~\cite{santos2025managing}.

Furthermore, by decomposing conflict detection, we can focus on the design choices and trade-offs of individual stages, while also enabling better categorization of existing works.

\section{Related Works}\label{sec:back}

The literature on conflict detection has primarily focused on \acs{O-RAN}, where 
\ac{AI} agents exist in the form of xApps and rApps, with explicit visibility 
into parameters and \acp{KPI}.
Early works focused on rule-based approaches for detecting conflicts based 
on the performance degradation of KPIs~\cite{wadud2025xapp, wadud2024qacm}.
These approaches set operating ranges for \ac{KPI} values and tracked 
parameter changes, characterizing conflicts as \ac{KPI} deviations from these 
ranges and attributing their cause to the most recent parameter change. 
These works assume all relationships between agents, parameters, and
\acp{KPI} are known in advance. As a result, they do not perform
interaction learning or graph reconstruction and only apply a rule-based
conflict identification. However, it is unclear how such relationships can be
determined, nor how this approach could be applied to networks
with different sets of agents, parameters, and \acp{KPI}.

To address this limitation, more recent works started 
learning interactions between parameters and \acp{KPI} to
reconstruct conflict graphs. For example, the work
of~\cite{del2025pacifista} generated statistical profiles of agents operating in
a sandbox environment, akin to a digital twin, learning interactions between
parameters and \acp{KPI} through \acp{ECDF}, and determined the 
relevant interactions for creating conflict graphs using the Kolmogorov–Smirnov 
distance, followed by a rule-based conflict
identification.
However, the effectiveness of the resulting conflict graph (and consequently, 
conflict detection) relies on the choice of an arbitrary set of tests and the 
fidelity of the sandbox in capturing realistic network behavior.

Other works have addressed the reliance on arbitrary testing within digital
twins by exploring data-driven techniques for learning how relationships between
parameters and \acp{KPI} emerge from data collected from the \ac{RAN}. 
For example, the work of~\cite{sharma2025towards} inferred cause and 
effect relationships between parameters and \acp{KPI} by adopting 
regression models for learning interactions between parameters and
\acp{KPI}, and leveraging \ac{SHAP} values with manually chosen thresholds
to create asymmetric conflict graphs.
This approach has the potential to enable root-cause
analysis for tracing the origin of a conflict. However, it is limited to
identifying indirect conflicts,
as it 
does not consider the relationships between agents with parameters and \acp{KPI}. 
Meanwhile, the work of~\cite{zolghadr2025learning}, adopted a \ac{GNN} for 
learning interactions between parameters and \acp{KPI} using 
correlation, and manually chosen thresholds to select relevant 
relationships to reconstruct conflict graphs, followed by a rule-based
conflict identification.
While this approach allows for the identification of conflicts with high accuracy,
\acp{GNN} are complex and computationally expensive, requiring
combinatorial exploration of interactions through repeated message-passing
operations~\cite{shao2024distributed}.

In this work, we propose a lightweight approach for learning
interactions between parameters and \acp{KPI}, and an autonomous approach 
for graph reconstructions. 
In the following two sections, we introduce \1 a two-tower encoder architecture 
inspired by the ranking problems encountered in recommender 
systems~\cite{yan2024ranktower}, that is simpler and more scalable than 
\ac{GNN}-based solutions, and \2 a sparsity-based approach for selecting 
relevant interactions~\cite{martins2016softmax}, that is more general 
than existing threshold-based solutions and can reconstruct graphs 
without manual fine-tuning to specific datasets or scenarios.

\section{Proposed Interaction Learning Method}\label{sec:tower}

To simplify training, reduce computational costs, and 
accelerate learning of interactions between parameters and 
\acp{KPI}, we propose a two-tower encoder architecture, as illustrated in Fig.~\ref{fig:tower_arch}. 
We adopt two independent encoders, known as towers, that project 
different types of input objects into a shared, lower-dimensional 
latent space, and then, we propose to calculate the relevance
of their interactions using scaled cosine similarity, akin to~\cite{li2022inttower}. 
In the following, we detail the formulation of our two-tower encoder architecture, 
and describe how we can reconstruct the score matrix in \eqref{eq:score_matrix}.

\subsection{Two-Tower Encoder Architecture}

We define a 
two-tower encoder architecture that independently encodes $\bm{X}_p$ 
and $\bm{X}_k$ into a shared latent space of dimension $H$:
\begin{subequations}
\begin{equation}
  \bm{Z}_p = f_p(\bm{X}_p, \theta_p) \in \mathbb{R}^{N_p \times H}, 
\end{equation}
\begin{equation}
  \bm{Z}_k = f_k(\bm{X}_k, \theta_k) \in \mathbb{R}^{N_k \times H},
\end{equation}
\end{subequations}
where $f_p(\cdot)$ and $f_k(\cdot)$ are learnable encoders,  
$\theta_p$ and $\theta_k$ denote the trainable weights of the 
parameter and \ac{KPI} encoders, respectively, and $\bm{Z}_p$ and 
$\bm{Z}_k$ represent the node embeddings of parameters and \acp{KPI} 
in the latent space, respectively.

\begin{figure}[t]
    \centering
    \includegraphics[width=0.99\columnwidth]{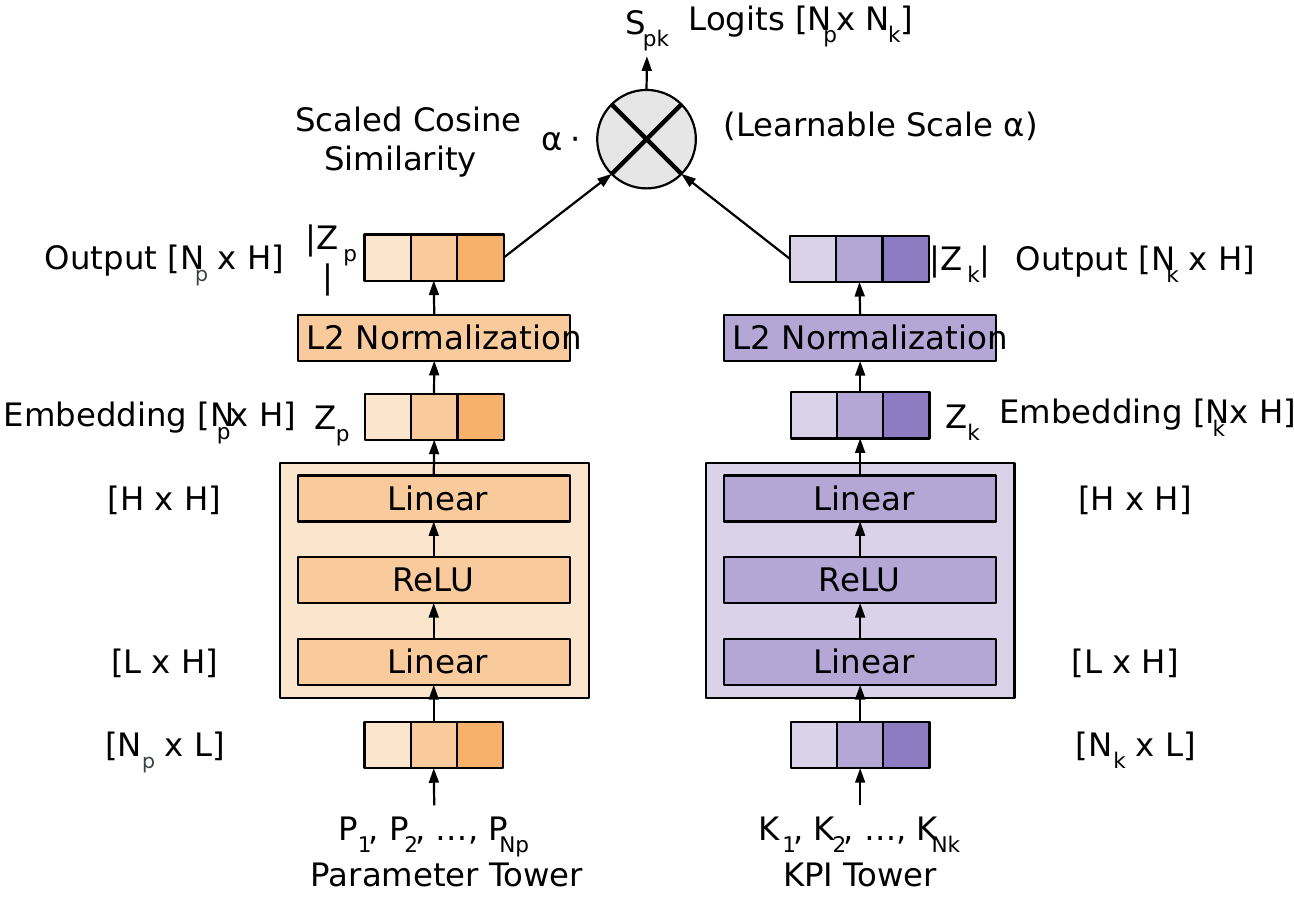}
    \vspace{-0.8em}
    \caption{Our two-tower encoder architecture, showing the 
      operations for independently encoding parameters and \ac{KPI} into the
      latent space. We calculate the similarity of their node embeddings 
      to infer interactions between parameters and \acp{KPI}.}
    \label{fig:tower_arch}
\end{figure}

It is worth noting that our two-tower encoder architecture does 
not depend on a specific encoder architecture. The encoders can be 
realized using different models depending on the nature of the training 
data and the objective of the learning task. 
In our current realization, each encoder comprises a lightweight 
feed-forward network composed of a linear layer followed by a 
\ac{ReLU} activation layer, and a final linear output layer, 
similar to~\cite{lebib2025two}, as illustrated in Fig.~\ref{fig:tower_arch}. 
Alternative models for the encoders may include, e.g., temporal models for 
capturing the dynamic evolution of the system~\cite{shen2022two}.
We detail the training of our two-tower encoder architecture later 
in Section~\ref{sub:train}.

To quantify the strength of the interactions between parameters and
\acp{KPI} in the latent space, we perform a scaled cosine similarity. 
To this end, we first apply a row-wise L2-normalization on the
node embeddings to ensure they have comparable scales, 
and prevent the embedding
magnitude from dominating the calculation of their similarity scores, i.e.:
\begin{subequations}
\begin{equation}
  |[\bm{Z}_{p}]_{i,:}| = \frac{[\bm{Z}_{p}]_{i,:}}{\|[\bm{Z}_{p}]_{i,:}\|_2}, \quad \forall i =
  1,\ldots,N_p,
\end{equation}
\begin{equation}
  |[\bm{Z}_{k}]_{j,:}| = \frac{[\bm{Z}_{k}]_{j,:}}{\|[\bm{Z}_{k}]_{j,:}\|_2}, \quad \forall j =
  1,\ldots,N_k,
\end{equation}
\label{eq:norm}
\end{subequations}
where $\|\cdot\|_2$ is the L2-norm operator, $[\cdot]_{i,:}$ is the $i$-th 
row of a matrix $\cdot$, while $|\cdot|$ denotes the normalized node embedding.

Then, we calculate the scaled cosine similarity between parameters and \acp{KPI}
via the matrix product of $|\bm{Z}_p|$ and $|\bm{Z}_k|$, quantifying the 
strength of their interaction based on their alignment in the latent space:
\begin{equation}
\bm{S}_{pk} = \alpha \, |{\bm{Z}_p}| |{\bm{Z}_k}|^\top \in \mathbb{R}^{N_p \times N_k},
\label{eq:sim}
\end{equation}

where $\bm{S}_{pk}$ is a partial score matrix containing similarity 
values between parameters and \acp{KPI}, and $\alpha \in \mathbb{R}_{+}$ is 
a learnable scale parameter for re-scaling the cosine 
similarity~\cite{yan2024ranktower}.

\subsection{Score Matrix Reconstruction}
\label{subsec:recon}

Once training is complete, $\bm{S}_{pk}$ enables the identification 
of \textit{indirect} conflicts between parameters and \acp{KPI}.
However, this partial matrix alone is insufficient for detecting
\textit{implicit} conflicts, which arise through complex chains of interactions
involving multiple parameters, \acp{KPI}, and \ac{AI} agents. Identifying such
conflicts also requires information about same-entity interactions among parameters 
and among \acp{KPI}, which are not directly observable from 
measurements of parameters and \acp{KPI}.
To address this issue, 
we train our two-tower architecture to learn cross-entity interactions between 
parameters and \acp{KPI}, from \eqref{eq:sim}, and subsequently leverage 
the learned node embeddings to infer same-entity interactions among 
parameters and among \acp{KPI}, and construct 
the score matrix in \eqref{eq:score_matrix}.

To accomplish this, first we concatenate the learned $\bm{Z}_p$ and $\bm{Z}_k$ 
after training
to form a unified 
embedding $\bm{Z}_{\texttt{all}}$:
\begin{equation}
\bm{Z}_{\texttt{all}} =
\begin{bmatrix}
\bm{Z}_p \\
\bm{Z}_k
\end{bmatrix}
\in \mathbb{R}^{(N_p + N_k) \times H}.
\end{equation}

Then, we apply the L2-norm on the unified embedding
$\bm{Z}_{\texttt{all}}$ to ensure it has a comparable scale regardless 
of the type of entity, analogous to \eqref{eq:norm}, creating 
$|\bm{Z}_{\texttt{all}}|$. Finally, we calculate the full score matrix 
$\bm{S}$ using a scaled cosine similarity:
\begin{equation}
\bm{S} = \alpha \, |\bm{Z}_{\texttt{all}}| |\bm{Z}_{\texttt{all}}|^\top \in \mathbb{R}^{(N_p + N_k) \times (N_p + N_k)},
\end{equation}

where $\alpha$ was learned during training from \eqref{eq:sim}, 
and $\bm{S}$ has the following block structure:
\begin{equation}
\bm{S} =
\begin{bmatrix}
\bm{S}_{pp} & \bm{S}_{pk} \\
\bm{S}_{kp} & \bm{S}_{kk}
\end{bmatrix},
\qquad
\bm{S}_{kp} = \bm{S}_{pk}^\top.
\end{equation}

The full score matrix $\bm{S}$ is symmetric and contains similarity scores for both 
cross-entity interactions between parameters and \ac{KPI} ($\bm{S}_{pk}$), 
as well as same-entity relationships among 
parameters ($\bm{S}_{pp} \in \mathbb{R}^{N_p \times N_p}$) and 
among \acp{KPI} ($\bm{S}_{kk} \in \mathbb{R}^{N_k \times N_k}$). 
While $\bm{S}_{pk}$ is learned from training on measurements between 
parameters and \acp{KPI}, the same-entity blocks 
$\bm{S}_{pp}$ and $\bm{S}_{kk}$ instead emerge after training by 
evaluating pairwise alignments among parameters and among 
\acp{KPI} in the shared latent space, respectively.

\section{Proposed Graph Reconstruction Method}\label{sec:thresh}

The score matrix $\bm{S}$ provides a soft representation of interactions
between entities,  
capturing information about the
behavior and state of the \ac{RAN}.
However, it does not provide the topological information required for 
identifying conflicts. 
To accomplish this, we reconstruct a conflict graph by mapping 
$\bm{S}$ into $\hat{\bm{A}}_{\texttt{learned}}$ with 
topological information that encodes potentially conflicting 
relationships. This process requires determining relevant interactions 
in $\bm{S}$ that constitute actual relationships, and binarizing 
them to obtain 
an\;adjacency\;matrix\;representing a conflict graph.

There are several techniques conventionally  used in machine learning tasks   
to determine relevant interactions based on the distribution of score
values. These include:
\1 selecting a reconstruction threshold $\tau$ and retaining all score values 
above it; \2 selecting a fixed number of $K$ relevant scores, and retaining  
the top-$K$ highest score values; or \3 choosing a representative $Q$ 
percentage of the score distribution, and retaining the top $Q\%$ of scores.
However, all these techniques share the same fundamental limitation: they require 
parameterization and manual fine-tuning on a given score distribution to
find appropriate values for $\tau$, $K$ or $Q$. As a result, conflict 
detection solutions using these methods are difficult to generalize 
across datasets and deployment 
scenarios~\cite{zolghadr2025learning, del2025pacifista}.

To address this limitation, we adopt
$\operatorname{sparsemax}$~\cite{martins2016softmax}, a data-driven, comparative 
normalization function. It maps real-valued scores to a probability distribution by  
projecting each row of $\bm{S}$ onto the probability simplex via 
an Euclidean projection, yielding a sparse representation that retains 
only strong interactions while weaker interactions collapse to zero.
As such, $\operatorname{sparsemax}$ provides a general and autonomous mechanism
for determining relevant interactions across datasets and 
deployment scenarios without manual fine-tuning.

First, we apply the $\operatorname{sparsemax}$ operator to $\bm{S}$ as follows:
\begin{equation}
  [\bm{P}]_{i,:} = \operatorname{sparsemax}([\bm{S}]_{i,:}), \quad \forall i = 1, \ldots, (N_p + N_k),
\end{equation}

where $\bm{P}\in [0,1]^{(N_p + N_k) \times (N_p + N_k)}$ denotes the 
resulting sparse probability matrix. 
Since $\operatorname{sparsemax}$ inherently collapses all weak interactions 
to exact zeros, we can retain all relevant interactions by binarizing 
$\bm{P}$ using a fixed threshold $\tau_{\text{sparse}} = 0$, yielding the learned 
adjacency matrix  $\hat{\bm{A}}_{\texttt{learned}}$:
\begin{equation}
  [\hat{\bm{A}}_{\texttt{learned}}]_{i,j} = \mathds{1}([\bm{P}]_{i,j} > 0), \quad i \neq j.
\end{equation}

Finally, we augment $\hat{\bm{A}}_{\texttt{learned}}$ with known information
about $\bm{A}_{\texttt{known}}$, as defined in~\eqref{eq:box}, yielding the 
full adjacency matrix $\hat{\bm{A}}$ that represents a conflict graph and 
supports the use of conflict identification methods~\cite{zolghadr2025learning}.

\section{Numerical Results}\label{sec:eval}

In this section, we validate our two-tower encoder architecture 
for interaction learning, and our sparsity-based approach for 
autonomous conflict graph reconstruction. First, we describe our 
dataset and training process. Then, we assess the performance 
of our two towers for learning interactions between parameters and \acp{KPI}, 
and its impact on conflict detection. 
Next, we assess the performance of $\operatorname{sparsemax}$ for
reconstructing graphs, and evaluate its impact on conflict detection.

\subsection{Dataset and Training}\label{sub:train}

To evaluate our proposed two-tower encoder architecture for interaction
learning, we adopted a dataset commonly used in the conflict detection
literature~\cite{zolghadr2025learning, wadud2024qacm}, 
based on a conflict model for cognitive autonomous networks
proposed in~\cite{banerjee2021toward}.
This dataset contains $L = 10,000$ samples of Gaussian-distributed
parameters and \acp{KPI}, capturing the evolution of $N_a = 4$ agents, 
$N_p = 7$ parameters, and $N_k = 4$ \acp{KPI}.
For information about the structure of the conflict model and how it defines the
interactions between agents, parameters, and \acp{KPI}, we refer the reader
to~\cite{zolghadr2025learning}.
To learn the interactions between the entities in our dataset, we designed two
separate encoders of parameters and \acp{KPI}, each tailored to 
the corresponding input dimensions. 
In addition, we used the Adam optimizer with a 
learning rate of 0.001~\cite{lebib2025two}, with the latent space dimension of 
$H = 16$, which performed the best in our evaluation.

We trained our model using the \ac{BCE} loss, 
to
minimize the difference between the learned similarities and the ground-truth
label matrix $\bm{Y} \in \{0,1\}^{N_p \times N_k}$. 

  The associated loss function is defined as: 
\begin{equation}
\mathcal{L}_{\mathrm{BCE}} =
\frac{1}{N_p N_k}
\sum_{i=1}^{N_p} \sum_{j=1}^{N_k}
\ell\!\left([\bm{S}_{pk}]_{i,j}, [\bm{Y}]_{i,j}\right),
\end{equation}

where $\ell(.)$ denotes the per-sample loss, given by 
$\ell(x, y) = \max(x, 0) - x\,y + \log\!\left(1 + e^{-|x|}\right)$.
After the training, we use the partial score matrix 
$\bm{S}_{pk}$ with learned interactions 
to reconstruct a complete score matrix $\bm{S}$, as detailed in
Section~\ref{subsec:recon}.

\begin{figure}[t]
    \centering
    \hfill
    \begin{subfigure}[b]{0.49\columnwidth}
        \centering
        \includegraphics[trim=0.8em 1em 1em 0.5em, clip, width=\textwidth]{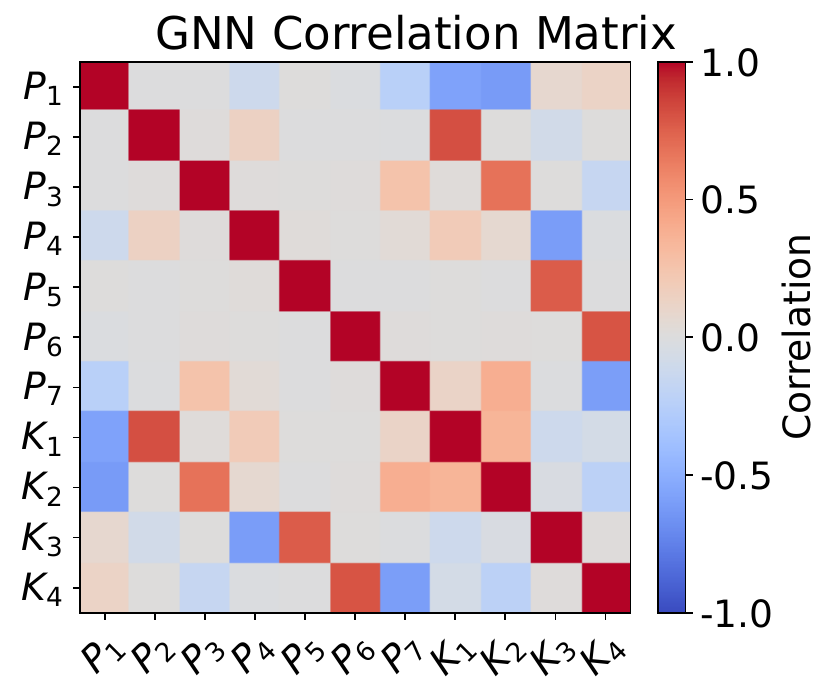}
    \end{subfigure}
    \hfill
    \begin{subfigure}[b]{0.49\columnwidth}
        \centering
        \includegraphics[trim=0.8em 1em 1em 0.5em, clip, width=\textwidth]{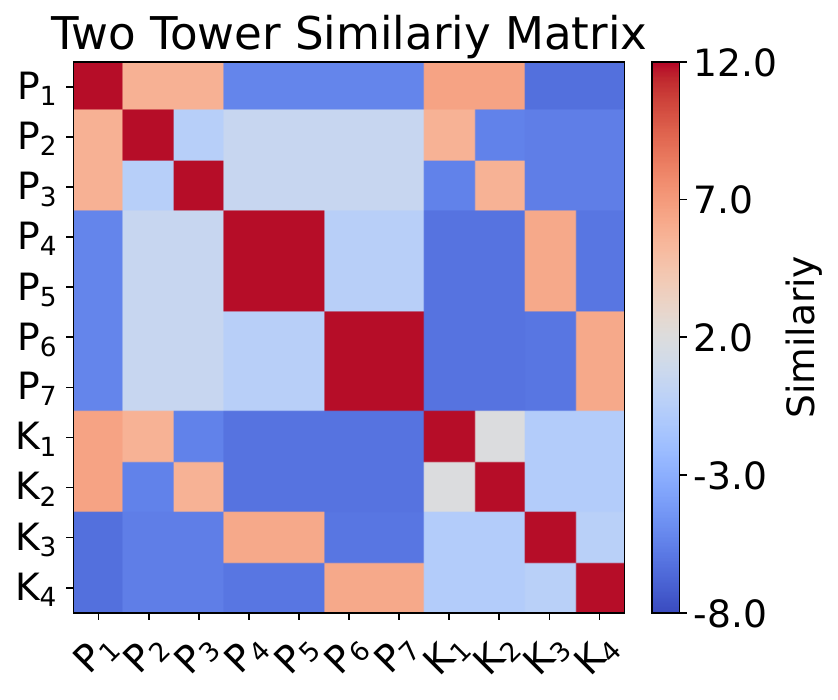}
    \end{subfigure}
    \hfill
    \vspace{-0.5em}
    \caption{Example score matrices capturing interactions between parameters
      and \acp{KPI}, highlighting the different metrics and scales associated with
    correlation (left) and similarity (right).}
    \label{fig:scores}
\end{figure}

\subsection{Two-tower Encoder Learning Performance}

\begin{figure}[t]
    \centering
    \includegraphics[width=.99\columnwidth]{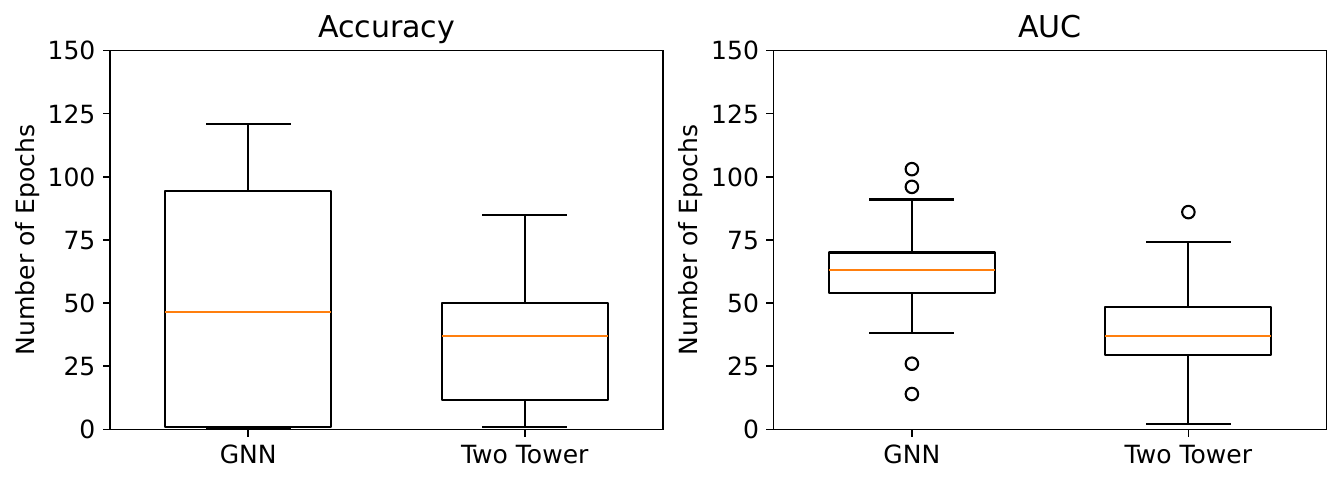}
    \vspace{-0.5em}
    \caption{Classification performance of different models in terms of
      accuracy and \acs{AUC}\acreset{AUC} across 100 independent training batches.}
    \label{fig:model_results}
\end{figure}

In this analysis, we evaluate our two-tower encoder
architecture's performance for learning interactions, 
allowing us to reconstruct conflict graphs and identify conflicts. 
We compared our two-tower model against the \ac{GNN} solution proposed 
in~\cite{zolghadr2025learning}. 
Fig.~\ref{fig:scores} presents an instance of the score matrix 
$\bm{S}$ output by the two models after training, where 
our two-tower model exhibits more distinct separation between 
classes than the \ac{GNN}, indicating that similarity is a 
more discriminative metric for learning interactions than correlation.
In Fig.~\ref{fig:model_results}, we assess the classification performance of the two models using training accuracy, capturing their ratio 
of correctly predicted interactions, and \ac{AUC}, capturing their 
ability to distinguish existing from non-existing interactions. The \ac{GNN}
requires a median of 47 and 63 training epochs to reach 80\% accuracy and
\ac{AUC}, respectively, whereas our model achieves the same
performance nearly twice as fast, at a median of 37 epochs for both metrics, due  
to the more discriminate class separation offered by the similarity 
scores, which simplifies the training of our two-tower model.

\begin{figure}[t]
    \centering
    \includegraphics[width=0.99\columnwidth]{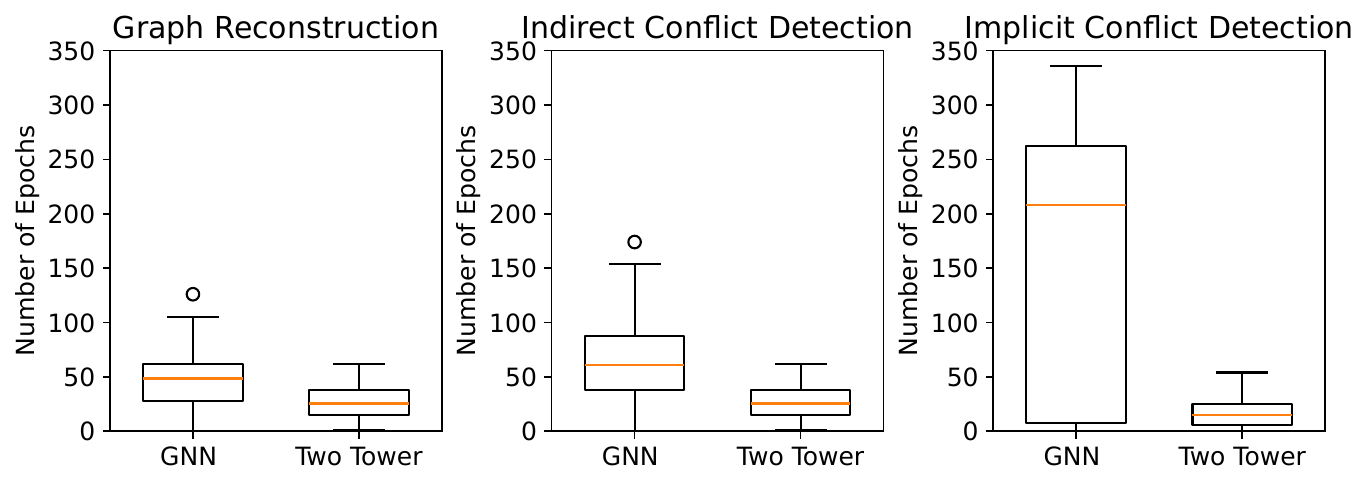}
    \vspace{-0.8em}
    \caption{Graph reconstruction and conflict identification performance using 
      the interactions learned from different models in terms of F1 score 
      across 100 independent training batches.}
    \label{fig:con_results}
\end{figure}

\subsection{Two-tower Encoder Detection Performance}
In this analysis, we assess the performance of the two models to 
express interactions for enabling the identification of indirect 
and implicit conflicts, using the F1 score.
To this end, we adopted the static threshold for score binarization 
and rule-based conflict identification approach proposed 
in~\cite{zolghadr2025learning}. Fig.~\ref{fig:con_results} shows the 
results of our measurements. The \ac{GNN} requires a median of 49 
epochs to enable the reconstruction of conflict graphs with perfect
performance ($F1 = 1.0$), whereas our two-tower model achieves the same twice as fast, at a median of 25 epochs. We can observe a 
similar behavior for indirect conflict detection, where the \ac{GNN} 
requires a median of 61 epochs, while our model reaches the same in 26 epochs. Moreover, the \ac{GNN} requires a median of 208 epochs to enable 
perfect detection performance, whereas our two-tower model attains 
the same after only 15 epochs, yielding nearly a $14\times$ 
improvement in training efficiency.
This is due to the different learning methods of the two models.
Our model directly calculates alignments in the 
latent space, whereas the \ac{GNN} progressively propagates information 
over message-passing iterations, slowing the 
emergence of complex interactions from implicit conflicts.
These results demonstrate that our two-tower encoder architecture
is more suited for learning complex interactions than existing \ac{GNN}-based solutions.

\begin{figure}[t]
    \centering
    \includegraphics[width=.99\columnwidth]{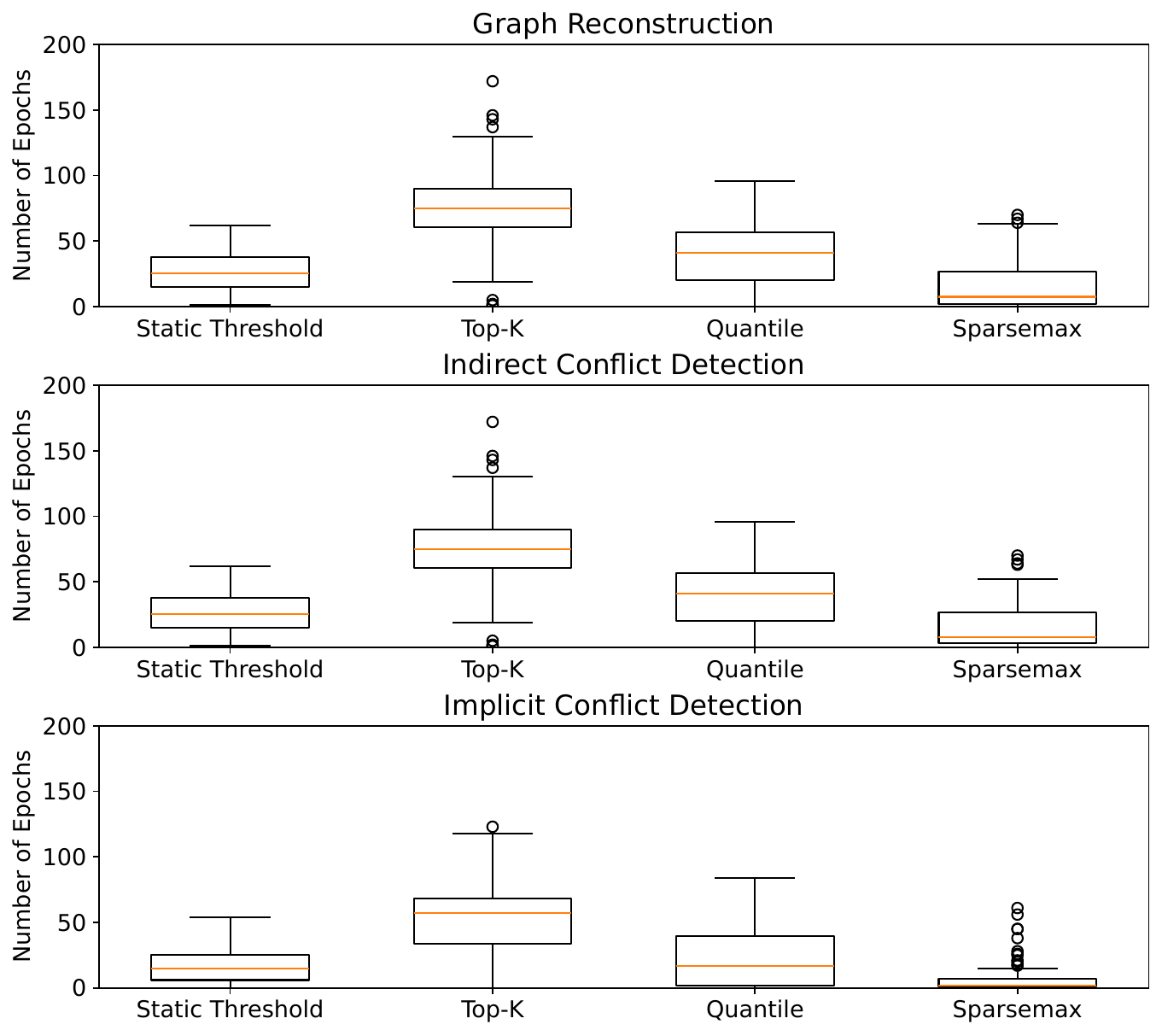}
    \vspace{-0.5em}
    \caption{Comparison of graph reconstruction methods in terms of training 
      efficiency required to achieve perfect 
      performance ($F1=1.0$) for conflict graph reconstruction and conflict 
      identification across 100 independent training batches.}
    \label{fig:tower_recon}
\end{figure}

\subsection{Autonomous Graph Reconstruction Performance}

In this analysis, we assess the performance of our proposed data-driven,
sparsity-based approach using $\operatorname{sparsemax}$ 
for autonomous graph reconstruction. 
We compared $\operatorname{sparsemax}$ against the techniques detailed in
Section~\ref{sec:thresh}, including a static threshold, top-$K$,
and statistical quantiles fine-tuned to our dataset, 
across 100 independent training batches. 
For the sake of brevity, we focus our evaluation 
on the interaction scores output by our two-tower model, 
which we have shown to offer significant advantages over existing 
\ac{GNN}-based solutions. 

For graph reconstruction,  shown in Fig.~\ref{fig:tower_recon} (top plot), 
we observe that the top-$K$ and quantile techniques require substantially 
stronger interactions resulting from longer training to achieve 
perfect reconstruction ($F1 = 1.0$),
being $\approx 2\text{--}3\times$ slower than the static 
threshold that achieves perfect reconstruction at a median of 26 epochs. 
In contrast, $\operatorname{sparsemax}$ achieves the same 
at a median of only 8 epochs, yielding a $4\times$
improvement in training efficiency. 
This is due to $\operatorname{sparsemax}$'s comparative normalization, which enforces competition among interaction scores, making subtle differences expressive 
early in training, enabling a faster graph reconstruction.

For conflict detection, we adopted the rule-based
identification approach proposed in~\cite{zolghadr2025learning}.
We observe a similar trend for indirect conflicts in 
Fig.~\ref{fig:tower_recon} (middle plot), where the top-$K$ and quantile 
techniques  
are $\approx 2\text{--}3\times$ slower than the static threshold 
that achieves perfect identification at a median of 26 epochs, 
while $\operatorname{sparsemax}$ achieves the same at a median of only 15 epochs, 
yielding $\approx 2 \times$ performance gains. 
For implicit conflicts, shown in Fig.~\ref{fig:tower_recon} (bottom plot), 
the top-$K$ and quantile are $1\text{--}5\times$ slower than the static 
threshold at a median of 15 epochs. Meanwhile, $\operatorname{sparsemax}$ 
reaches the same at a median of only 2 epochs, delivering nearly an $8\times$
performance gain.
These results demonstrate that $\operatorname{sparsemax}$ is a more effective 
mechanism for expressing relevant interactions for graph reconstruction, 
outperforming other methods while remaining fully autonomous and requiring no
manual fine-tuning.

\section{Conclusions}\label{sec:conc}

In this paper, we introduced a systematic framework for conflict detection in
\ac{AI}-native mobile networks, decomposing the problem into three stages:
interaction learning, graph reconstruction, and conflict identification. We
formulated interaction learning as a ranking problem between parameters and
\acp{KPI} and proposed a lightweight two-tower encoder architecture for 
efficiently learning these interactions. In addition, we introduced a data-driven,
sparsity-based approach for selecting relevant interactions, enabling autonomous
conflict graph reconstruction without manual fine-tuning for specific datasets or 
scenarios. 
We showed that our proposed solution outperforms state-of-the-art methods 
across a variety of conflict detection tasks, from conflict graph reconstruction to the identification of different types of conflicts. 
In future work, we plan to 
explore alternative conflict identification approaches, 
investigate temporal models to capture the dynamic evolution of the \ac{RAN}, and 
deploy our proposed solutions in a \ac{AI}-native network environment.

\section*{Acknowledgements}

Work supported through the INL Laboratory Directed Research \& Development
(LDRD) Program under DOE Idaho Operations Office Contract DE-AC07-05ID14517.
This research was also partially funded by the NSF IUCRC WISPER, under NSF Award
Nos. 2412872, 2413009, and 2413168, and by the NSF US-Ireland R\&D Partnership 
program under NSF Award No. 2421362.
This work also received support from the Horizon Europe SNS JU
program under grant No. 101139194 (6G-XCEL).
The research leading to this paper also received support from the Commonwealth
Cyber Initiative, an investment in the advancement of cyber R\&D, innovation, and workforce development. 
For more information, visit: \url{www.cyberinitiative.org}.

\bibliographystyle{IEEEtran}
\bibliography{IEEEabrv,bibliography}


\end{document}